\documentstyle[12pt,amsfonts]{article}
\parskip=\medskipamount

\newcommand {\cA}{{\cal A}}

\newcommand {\cF}{{\cal F}}

\newcommand {\cJ}{{\cal J}}

\newcommand {\cL}{{\cal L}}
\newcommand {\cM}{{\cal M}}

\def\a{\alpha}
\def\b{\beta}

\def\d{\delta}

\def\g{\gamma}
\def\G{\Gamma}

\def\k{\kappa}

\def\o{\omega}

\def\q{\theta}

\def\s{\sigma}

\def\D{\Delta}
\def\F{\Phi}
\def\J{\Psi}
\def\L{\Lambda}

\def\S{\Sigma}
\def\U{\Upsilon}

\newcommand{\ad}{{\dot{\alpha}}}                           
\newcommand{\bd}{{\dot{\beta}}}                            
\newcommand{\ve}{\varepsilon}                            

\newcommand{\pa}{\partial}                           
\newcommand{\hf}{\frac12}

\newcommand{\sect}[1]{\setcounter{equation}{0}\section{#1}}

\newcommand{\be}{\begin{equation}}
\newcommand{\ee}{\end{equation}}
\newcommand{\bea}{\begin{eqnarray}}
\newcommand{\eea}{\end{eqnarray}}
\newcommand{\non}{\nonumber}

\def\Bar#1{\overline{#1}}
\newcommand{\beq}{\begin{equation}}
\newcommand{\eeq}{\end{equation}}
\def\eqalign#1{\,\vcenter{\openup2\jot \caja
        \ialign{\strut \hfil$\displaystyle{##}$&$
        \displaystyle{{}##}$\hfil\crcr#1\crcr}}\,}
\def\caja{\mathsurround=0pt}
\def\un#1{\underline {#1}}
\def\fracm#1#2{\hbox{\large{${\frac{{#1}}{{#2}}}$}}}

\font\ro=cmsy10       
\def\kcr{{\hbox{\ro \char'170}}}    
\def\ktl{{\hbox{\ro \char'170}}}  
\def\ktr{{\hbox{\ro \char'170}}}  
\def\kbl{{\hbox{\ro \char'170}}}  
\def\kbr{{\hbox{\ro \char'170}}}  

\def\endtitle{\end{quotation}\newpage}     

\def\headpic{           
  \indent
  \setlength{\unitlength}{.4mm}
  \thinlines
  \par
  \begin{picture}(29,16)
  \put(165,16){\line(1,0){4}}
  \put(170,16){\line(1,0){4}}
  \put(180,16){\line(1,0){4}}
  \put(175,0){\line(1,0){4}}
  \put(180,0){\line(1,0){4}}
  \put(185,0){\line(1,0){4}}
  \put(169,0){\line(0,1){16}}
  \put(170,0){\line(0,1){16}}
  \put(179,0){\line(0,1){16}}
  \put(180,0){\line(0,1){16}}
  \put(184,0){\line(0,1){16}}
  \put(185,0){\line(0,1){16}}
  \put(169,16){\oval(8,32)[bl]}
  \put(170,16){\oval(8,32)[br]}
  \put(179,0){\oval(8,32)[tl]}
  \put(185,0){\oval(8,32)[tr]}
  \end{picture}
  \par\vskip-6.5mm
  \thicklines}

\def\border{           
  \setlength{\unitlength}{1mm}
  \newcount\xco
  \newcount\yco
  \xco=-21
  \yco=12
  \begin{picture}(140,0)
  \put(\xco,\yco){$\ktl$}
  \advance\yco by-1
  {\loop
  \put(\xco,\yco){$\kcr$}
  \advance\yco by-2
  \ifnum\yco>-240
  \repeat
  \put(\xco,\yco){$\kbl$}}
  \xco=158
  \yco=12
  \put(\xco,\yco){$\ktr$}
  \advance\yco by-1
  {\loop
  \put(\xco,\yco){$\kcr$}
  \advance\yco by-2
  \ifnum\yco>-240
  \repeat
  \put(\xco,\yco){$\kbr$}}
  \put(-20,13){\tiny University of Maryland Elementary Particle
Physics University of Maryland Elementary Particle Physics University of
Maryland Elementary Particle Physics}
  \put(-20,-241.5){\tiny University of Maryland Elementary
Particle Physics University of Maryland Elementary Particle Physics
University of Maryland Elementary Particle Physics}
  \end{picture}
  \par\vskip-8mm}

\begin{document}
\border\headpic {\hbox to\hsize{October 1998 \hfill {UMDEPP 98-133}}}
\par
\setlength{\oddsidemargin}{0.3in}
\setlength{\evensidemargin}{-0.3in}
\begin{center}
\vglue .03in
{\large\bf The CNM-Hypermultiplet Nexus}\footnote{Supported in
part by NSF Grant PHY-98-02551 and Deutsche Forschungsgemeinschaft.}
\\[.4in]
S. James Gates, Jr. \\
{\it Department of Physics,
University of Maryland at College Park \\
College Park, MD 20742-4111, USA}\\
{\tt gates@bouchet.physics.umd.edu}\\
[0.2in]
and \\[0.2in]
Sergei M. Kuzenko \footnote{On leave from: Department of Physics,
Tomsk State University, Tomsk 634050, Russia}\\
{\it Sektion Physik, Universit\"at M\"unchen\\
Theresienstr. 37, D-80333 M\"unchen, Germany}\\
{\tt sergei.kuzenko@physik.uni-muenchen.de
} \\[1.4in]

{\bf ABSTRACT}\\[.002in]
\end{center}
\begin{quotation}
{We consider additional properties of CNM (chiral-nonminimal) models. 
We show how 4D, $N$ = 2 nonlinear $\s$-models can be described {\it {solely}} 
in terms of $N$ = 1 superfield CNM doublets. These actions are described 
by a K\" ahler potential together with an infinite number (in the general 
case) of terms involving its successively higher derivatives. We briefly 
discuss how $N$ = 2 supersymmetric extension of the previously proposed 
$N$ = 1 CNM low-energy QCD effective action can be achieved.}\endtitle

\sect{Introduction}

~~~~The nonminimal scalar multiplet \cite{GaS}, long ignored, is contained 
in a complex linear superfield and like the chiral scalar multiplet only
describes physical fields of spin-0 and spin-1/2. Previously \cite{DeoG}, it 
was suggested that using chiral-nonminimal (CNM) doublets leads to a natural 
introduction of Dirac particles in 4D, $N$ = 1 supersymmetrical theories. A 
hallmark of CNM models is that the left-handed components of Dirac spinors 
appear in {\it {distinct}} supersymmetry representations from the corresponding 
right-handed components.  In fact, CNM  doublets can never be used to describe 
Majorana particles, unlike theories  with solely chiral superfields.  Recently 
nonminimal scalar multiplets  have appeared in a number of investigations.  It
has   been proposed to construct 4D, $N$ = 1 supersymmetric extensions of ``soft 
pion physics'' \cite{SM} by use of CNM doublets \cite{G1}. Similarly CNM
doublets  have been used to discuss 2D, $N$ = 2,4 models  (obtainable simply 
from dimensional reduction) \cite{PRSZ}. In conjunction with performing 
explicit calculations \cite{GRR} within the context of matter coupled to 
4D, $N$ = 2 supersymmetric Yang-Mills theory and the Seiberg-Witten conjecture 
\cite{SW}, CNM doublets have emerged.  The CNM doublets also underlie the 
parallel development of these theories via harmonic superspace \cite{BBIKO} and 
their role has been recently clarified \cite{SK} within the harmonic superspace 
approach \cite{GIKOS,2SYMH} to the manifest realization of 4D, $N$ = 2 
supersymmetry.  The work of ref.~\cite{SK} also provides a simple answer to 
why only Dirac particles (not Majorana particles) arise in CNM models.  The
`q-hypermultiplets'  of harmonic superspace can only describe  Dirac particles
and  breaking the $N$ = 2 supersymmetry to $N$ = 1 preserves this result. The 
dynamical $N$ = 1 superfield content of `q-hypermultiplets' consists of CNM
doublets.

There is a very old result similar to the observations above. In the nomenclature 
of harmonic superspace, there is a second hypermultiplet, called the
``$\o$-hypermultiplet'' \cite{GIKOS,2SYMH}.  This manifest 4D, $N$ = 2 
representation is closely related to the ``relaxed hypermultiplet'' \cite{HST}. 
It is known that the relaxed  hypermultiplet can also be decomposed into $N$ = 1
superfields.  When  this is done \cite{GGRS}, the dynamical $N$ = 1 superfields 
that emerge  also involves a CNM doublet. The same is true for higher relaxed 
hypermultiplets \cite{YS,GIOmost}.  Moreover, any consistent projective
truncation of the $\o$-hypermultiplet, which can be carried out along the line
described in \cite{SK}, leads to a matter multiplet involving a CNM doublet. We 
can thus make the following observation.

${~~~~}$ {\it {Every known formalism which possesses manifest off-shell
4D, $N$ = 2 \newline ${~~~~}$ supersymmetry and describes minimal
irreducible hypermultiplet represent- \newline ${~~~~}$ ations, decomposes
into CNM doublets under $N$ = 1 superfield reduction.}}

At this point, it is worth mentioning other historical antecedents. 
A 4D, $N$ = 1 superfield action involving the nonminimal multiplet and 
arbitrary numbers of nondynamical auxiliary superfields was proposed 
in \cite{DeoG}.  Even more closely related, expansions of the type above, 
but for {\it (in){finite}} series in $w$ were first suggested in \cite{KLR}.
This work was also the first to propose projective expansions in terms 
of $N$ = 1 superfields as a way to realize $N$ = 2 supersymmetry. However, 
this discussion did not extend to nonlinear $\s$-models.

With these new developments, we believe that CNM models will play an 
increasingly important role in our understanding of supersymmetry.  In 
this letter, we wish to add to this growing inventory of applications 
of CNM models by presenting more such results. 

These may be viewed as a simplification in 4D, $N$ = 2 nonlinear
$\s$-models.  It has long been asserted that such models must parametrize
hyper-K\" ahler manifolds. Such theories have been discussed completely
within the context of the harmonic superspace approach and as well within
the confines of projective superspace \cite{KLR}. We shall show that
a large class of N = 2 nonlinear $\s$-models in the projective
approach, are in fact completely expressible as CNM nonlinear $\s$-models.

We also briefly discuss how our observations provide the starting point 
for the construction of an $N$ = 2 supersymmetric extension to the 
previous work \cite{G1} of the class of $N$ = 1 supersymmetric actions
based on the idea of using CNM doublets to extend  non-supersymmetric
chiral perturbation theory (``soft pion physics'') \cite{SM}. 

\sect{$N$ = 1 CNM Formulation of 4D, $N$ = 2 Non-linear
\protect\newline $\sigma$-model Actions}

~~~~Presently, there are only two known formalisms that make manifest
4D, $N$ = 2 supersymmetry; (a.) the harmonic superspace approach \cite{
GIKOS} and (b.) the projective superspace approach \cite{KLR}.  Recently it was
explicitly  demonstrated \cite{SK} how these two approaches (not surprisingly) 
are related.  Any theory that is formulated in projective superspace can 
be re-written as an equivalent theory in harmonic superspace.  More 
unexpectedly, however, is that this recent explicit demonstration makes 
it clear that the $N$ = 1 superspace decomposition  of the ``q-hypermultiplet' 
of the harmonic approach, {\it {necessarily}} leads to the appearance of 
CNM doublets.  From this view, the appearance of  4D, $N$ = 1 CNM models 
is closely related to the potential realization of manifest 4D, $N$ = 2
supersymmetry. In our prior work \cite{G1} on CNM models, the issue of 
4D, $N$ = 2 supersymmetry has played no part in our considerations. 
Accordingly the progress of the two lines of research of \cite{G1} and 
\cite{GRR,KLR} developed independently. However, with the observation
of \cite{SK} it is clear that these two can coalesce around a nexus.
In particular the projective representations first suggested in \cite{KLR}
and more completely proffered in \cite{GRR} provide a convenient set of 
tools for finding where in the model space of CNM actions it becomes
possible to describe 4D, $N$ = 2 hypermultiplet nonlinear $\s$-models.
Although the necessary presence of CNM doublets arising from manifest 
4D, $N$ = 2 theories is surprising, the close relation between 4D, 
$N$ = 1 CNM models and 4D, $N$ = 2 supersymmetry is not. In fact, it was
noted \cite{DeoG} long ago that the CNM model defined on a ``flat'' K\"
ahler metric automatically possesses an on-shell $N$ = 2 supersymmetry.
The use of projective superspace techniques (the minimal reduction
of harmonic superspace) permits the realization of this $N$ = 2
supersymmetry beyond the limit of ``flat'' K\" ahler metrics in 4D, $N$ 
= 1 CNM models.

\subsection{Lifting $\sigma$-models:  $N = 1 ~ \to ~ N = 2$}
~~~~As is well known, the $N$ = 1 superfield action
\beq
{\cal S}_{\s} ~=~  \int d^8 z ~ K(\Phi^{I},
\, {\bar \Phi}{}^{\bar{I}}) ~~~, \label{nact4}
\eeq
leads to the component result (in the notation of `{\it Superspace}' \cite{GGRS})
\beq
\eqalign{
{{\cal S}}_{\s} \,=\, \int d^4 x \, \Big[  &- \fracm 12 g_{I {\bar J}
}\, ({\pa}^{\underline a} {\Bar A}{}^{\bar{J}} \, )  ({\pa}_{\underline a}
A{}^{I} \, ) \,-\, i g_{I {\bar J} } \, {\Bar \psi}{}^{\ad}{}^{\bar
J} {\cal D}_{\underline a} {\psi}^{\a}{}^{I} \,+\, \fracm 14 \, 
{\psi}^{\a}{}^{I}{\psi}_{\a}{}^{J} {\Bar \psi}{}^{\ad}{}^{\bar K}{\Bar
\psi}{}_{\ad}{}^{\bar L} R_{I  \bar{K} J \bar{L} } \, \Big] ~~~,
 \label{action2} } \eeq
after elimination of the auxiliary $F$-fields by their equations of motion.

We consider the following 4D, $N$ = 2 nonlinear $\s$-model given in $N$ = 1
superspace
\bea
S_{\s}[\U, \breve{\U}] \, &=& \, \int {\rm d}^8 z \,  \Big[ \, \,
\frac{1}{2\pi {\rm i}} \, \oint \frac{{\rm d}w}{w} \,  \, K \big( \U^I ,
\breve{\U}^{\bar{I}}  \big) ~~ \Big]  ~~~.
\label{nact} \eea
For $\U$ we have
\bea
\U^I  &=& \sum_{n=0}^{\infty}  \, \U_n^I (z) w^n ~=~ 
\F^I(z) + w \S^I(z) + O(w^2) ~~~,\non \\
\breve{\U}{}^{\bar{I}}  &=& \sum_{n=0}^{\infty}  \, {\bar \U}_n^{\bar{I}}
(z) (\fracm {-1}{w})^n ~=~ {\bar \F}{}^{\bar{I}}(z)  - \fracm 1{w}
{\bar \S}{}^{\bar{I}}(z)  + O((\frac{1}{w})^2)~~~,
\label{exp}
\eea
with $\F$ being chiral, $\S$ being complex linear, and the remaining
component superfields being unconstrained complex superfields. 
\be
{\Bar D}_\ad \F =0 ~~~, ~~~ ({\Bar D})^2 \S = 0 ~~~.
\ee
The  expansions in (\ref{exp}) describe ``polar'' multiplets in the nomenclature 
of \cite{GRR}. If we terminate the series in (\ref{exp}) at a finite value,
say $p$, and place a reality constraint on the highest $N$ = 1 superfield in
the expansion, this maintains $N$ = 2 supersymmetry.  Such multiplets, first
presented in \cite{KLT}, are called $O(p)$ multiplets in the nomenclature 
of \cite{GRR}.

The action possesses a linearly realized $N=2$ supersymmetry  (see Appendix A)
even if one replaces $K(\U, \breve{\U})$ by a more general $w$-dependent 
Lagrangian $K(\U, \breve{\U}, w)$ of the form
\be
K(\U, \breve{\U}, w) ~=~ \sum_{n} K_n(\U, \breve{\U}) w^n~~~,~~~ 
\Bar{K}_n ~=~ (-1)^n K_{-n}~~~.
\label{w-deplag}
\ee
A specific feature of the above action is that it remains invariant under
rigid U(1)  transformations
\be
\U(w) ~~ \longrightarrow ~~ \U({\rm e}^{{\rm i} \a} w) 
\quad \Longleftrightarrow \quad 
\U_n(z) ~~ \longrightarrow ~~ {\rm e}^{{\rm i} n \a} \U_n(z) 
~~~.
\label{rfiber}
\ee
Such transformations are compatible with the conjugation (\ref{pconjugation})
and can be treated as chiral rotations of those fermionic coordinates of the
$N=2$ superspace, which are eliminated in the $N=1$ approach. 

The $N=2$ sigma-model introduced respects all the geometric features of its $N=1$ 
predecessor in (\ref{nact4}). The K\"ahler invariance of (\ref{nact4})
\be
K(\F, \bar \F) \quad \longrightarrow \quad K(\F, \bar \F) ~+~ \Big( \L(\F)
\,+\,  {\Bar \L} (\bar \F) \Big)
\ee
turns into 
\be
K(\U, \breve{\U})  \quad \longrightarrow \quad K(\U, \breve{\U}) ~+~
\Big(\L(\U) \,+\, {\Bar \L} (\breve{\U} ) \Big)
\ee
for the model (\ref{nact}). A holomorphic reparametrization $A^I ~\rightarrow~ 
f^I \big( A \big)$ of the K\"ahler manifold has the following counterparts
\bea 
\F^I  \quad & \longrightarrow  & \quad f^I \big( \F \big) \non \\ 
\U^I (w) \quad & \longrightarrow & \quad f^I \big (\U(w) \big)
\eea
in the $N=1$ and $N=2$ cases, respectively. Therefore, the physical superfields
of the $N=2$ theory
\be
\U^I (w)\Big|_{w=0} ~=~ \F^I ~~~,~~~ \frac{ {\rm d} \U^I (w) }{ {\rm d} 
w} \Big|_{w=0} ~=~ \S^I ~~~,
\label{geo3} 
\ee
should be regarded, respectively,  as a coordinate of the K\" ahler manifold and 
a tangent vector at point $\F$ of the same manifold \cite{G1,SK}. That is why the variables
$(\F^I, \S^J)$ parametrize the tangent  bundle of the K\"ahler manifold.  Thus, 
the ad hoc geometrical properties assigned in \cite{G1} can be derived from 
projective superspace.  Our discussion to this point seems to have the 
following rather surprising implication.

${~~~~}$ {\it {Every 4D, N = 1 supersymmetric nonlinear $\s$-model
described by ({\ref {nact4}}) 
\newline ${~~~~}$ possesses a minimal extension to a 4D, N = 2
supersymmetric nonlinear \newline
${~~~~}$ $\s$-model described by ({\ref {nact}}) .}}

Let us represent  $\U (w)$ in the form :
\be
\U(w) ~=~ [\, \F \,+\, \S w \,]  ~+~ {\cal A} (w) ~~~,
\label{ansatz1}
\ee
where the quantity ${\cal A} (w)$ contains all the auxiliary superfields
at quadratic and higher powers in $w$. Since off-shell the dynamical
superfields $\F$, $\S$ and the auxiliary ones $\cA(w)$ are functionally
independent, we can write (\ref{nact}) in the form
\bea
S_{\s}[\U, \breve{\U}] &=& \int {\rm d}^8 z \,  \Big\{ \, \, \frac{1}{2\pi
{\rm i}} \, \oint \frac{{\rm d}w}{w} \, \, {\rm e}^{[~ {\cal A} \pa \, + \,
{\breve{\cal A}}  {\Bar \pa} ~ ] }~  K \big( \F \,+\, \S w , {\bar \F} \,-
\, \fracm 1{w}  {\bar \S} \big)  ~ \Big\} ~~~, \non  \\
&=& \int {\rm d}^8 z \, \Big\{  \, \, \frac{1}{2\pi {\rm i}} \, \oint 
\frac{{\rm d}w}{w} \, \, {\rm e}^{[~ {\cal A} \pa \, + \, {\breve{\cal A}}  
{\Bar \pa} ~ ] }~  \exp[\, w \S \pa - \fracm 1{w} \bar \S {\Bar
\pa} \, ]~   K \big( \F , {\bar \F} \, \big)  ~ \Big\} ~~~, {~~}
\label{nact2} \eea
where we have introduced the notation $\pa \equiv \pa/ \pa \Phi$ and 
${\Bar \pa} \equiv \pa/ \pa {\Bar \Phi}$.  This last form of writing 
the action  makes clear a number of features.  For example if $K ( \F ,
{\bar \F})$  is a finite polynomial of order $p$, then the highest power
to  which the auxiliary superfields appear is $p$.  

Given the K\"ahler potential $K(\U{}^I , \breve{\U}{}^{\bar I} )$, let us 
introduce the notation
\be
K_I (\U, \breve{\U}) ~=~ \frac{\pa~\,~~}{\pa \U^I} K (\U, \breve{\U}) 
\qquad  ~~~,
\ee
and suppose that $\U_{\star} (w)$ is a solution to the equations of motion
for the auxiliary superfields. These  equations read
\be
\frac{1}{2\pi {\rm i}} \oint \frac{{\rm d} w}{w} \,w^n \, K_I (\U_{\star},
\breve{\U}_{\star})  ~=~ 0 ~~~, \qquad n \geq 2 ~~~ .
\label{int}
\ee
The full mass shell is  described by adding the equation of motion for $\S$
\be
{\Bar D}_\ad \oint \frac{{\rm d}w}{w}\, w \, K_I \left( \U_\star , 
\breve{\U}_\star \right) ~=~ 0 ~~~,
\label{seq}
\ee
and the equation of motion for $\F$
\be
({\Bar D})^2 \oint \frac{{\rm d}w}{w}\,  K_I \left( \U_\star , \breve{\U}_\star 
\right) ~=~ 0 ~~~.~~~
\label{feq}
\ee

Our goal is to eliminate the infinite number of $N$ = 1 auxiliary superfields
 and reveal the explicit presence of a CNM model. Although it is relatively
simple to write the equations of motion for the auxiliary superfields
(\ref{int}), the actual elimination of the infinite  number of auxiliary
fields is problematical in the general case. In fact  to this point, we have
only a perturbative approach and special exact solutions  as  examples.  We will now
turn to the discussion of these. 

\subsection{Perturbative Elimination of Auxiliary Superfields}
~~~~Upon writing $ K = \U \breve{\U}$, we see that the action describes a free, 
massless CNM doublet $(\Phi^I, \, \S^I)$  together with an infinite number 
of $N$ = 1 auxiliary superfields whose purpose is to linearly realize
an $N$ = 2 supersymmetry.  Let us specialize the K\" ahler potential in
(\ref{nact}) to the form (field indices are suppressed)
\be 
K(\U, \breve{\U} ) ~=~ \U \breve{\U} ~+~ \ve {\cal K}(\U, \breve{\U})
~~~, \label{spec}
\ee
where $\ve$ is a small parameter, and ${\cal K}(\U, \breve{\U})$ is an
analytic function of $\U$ and $\breve{\U}$ (possessing a convergent Taylor 
series in an open vicinity of each point). The partial derivative of 
$K(\U, \breve{\U})$ with respect to $\U$ reads
\be
K_{\U} (\U, \breve{\U}) ~=~ \breve{\U} ~+~ \ve {\cal K}_{\U} (\U,
\breve{\U})~~~.
\ee
We continue by expanding ${\cal A}_\star$ (assumed to be a solution
to the equations (\ref{int}))
as a power series in $\ve$
\be  ~~~ {\cal A}_\star (w) ~\equiv~ \sum_{p =1}^{\infty} \ve^p \D \U_p (w)~~~, 
~~~ \D \U_p (w) ~=~ \sum_{m=2}^{\infty} \U_{p, m} w^m ~~~.
\label{ansatz2}
\ee
Inserting eqs. (\ref{ansatz1}, \ref{ansatz2}) into (\ref{int}), we obtain
\bea
(-1)^n \sum_{p=1}^{\infty} \ve^p  \bar{\U}_{p,n} ~+~ \ve (\frac{1}{2\pi 
{\rm i}}) \oint \frac{ {\rm d} w}{w} w^n {\cal K}_{\U} \left( \U_{\star}, 
\breve{\U}_{\star}  \right) ~=~ 0 ~~~, ~~~n>1~~~.
\label{ase2}
\eea
These equations allow us to solve for $\D \U_p (w)$ iteratively. To linear 
in $\ve$ order we obtain
\be
(-1)^n \bar{\U}_{1,n} +
\frac{1}{2\pi {\rm i}} \oint \frac{ {\rm d}
w}{w} w^n {\cal K}_{\U}  (\F + \S w ,~ \bar \F - \bar \S /{w} ) ~=~ 0
~~~~~, ~~~n>1 ~~~.
\label{ase3} \ee
Having determined $\D \U_1 (w)$, equations (\ref{ase2},\ref{ase3}) allow us
to  derive $\D \U_2 (w)$, 
\bea
(-1)^n \bar{\U}_{2,n}   &+& 
\frac{1}{2\pi {\rm i}} \oint \frac{ {\rm d} w}{w} w^n 
\left\{ \D \U_1 (w) \pa + \D \breve{\U}_1(w) \Bar{\pa} \right\} \non \\
& \times &
{\cal K}_{\U}  (\F + \S w ,~ \bar \F - \bar \S /{w} ) ~=~ 0
~~~~~, ~~~n>1 ~~~,
\eea
and on and on (at least in principle) to all orders.
The result of this process is that the auxiliary superfields are replaced
by their on-shell values,  i.e. ${\cal A}(w) \to {\cal A}_\star (w; \F, \bar \F, 
\S, \bar \S)$. We thus have a proof that there exists a
large class of 4D, $N$ = 2 nonlinear $\s$-models that permit the
``perturbatively  elimination'' of an {\it {infinite}} number of $N$ = 1
auxiliary fields leaving behind a dynamical system described solely in terms
of the CNM doublet $(\F , \, \S)$.

The above method can be applied for eliminating the auxiliary superfields
in more general $N=2$ models described by $w$-dependent potentials
(\ref{w-deplag}). A particular class of $w$-dependent Lagrangians on 
which we plan further study takes the form (below $\breve{w} = - 1/w$)
\be
K ~\equiv~ K ( \U , \breve{\U} ) ~+~ [ ~{\cal J} ( \U, \, {\breve w} 
) ~+~ {\Bar \cJ} (\breve{\U}, w) ~]  ~~~,  \label{HD}
\eeq
Provided $K( \U, \breve{\U})$ is $R$-invariant, $K( {\rm e}^{{\rm i} 
\varphi } \U, {\rm e}^{-{\rm i} \varphi } \breve{\U}) = K\left( \U, 
\breve{\U}\right)$, the action may be invariant  under generalized 
U(1) transformations (\ref{rfiber})
\be
\U(w) ~~ \longrightarrow ~~  {\rm e}^{{\rm i} \phi} \U({\rm e}^{{\rm 
i} \a} w) ~~~,
\ee 
where the parameter $\phi$ is determined by the holomorphic potential  
${\cal J} ( \U, \breve{w})$.

Finally, we observe that although our method of proof required that the
parameter $\ve$ be small in some sense, depending on the convergence
properties of the iterative solution, it may be possible to drop this
requirement and simply regard $\ve$ as a device for deriving the pure
CNM formulation of the theory.

\subsection{Special Exact Solutions}
~~~~Below we shall describe a technique to solve equations (\ref{int}) for a 
large family of K\"ahler manifolds including the complex projective spaces 
${\mathbb C}{\rm P}^n$ and the complex Grassmann spaces ${\rm G}_{m,n}$.

Let us choose the two-sphere ${\mathbb C}{\rm P}^1 ={\mathbb C} \, \cup
\,\{\infty \}$ in the role of our target K\"ahler manifold. Introducing a
complex local coordinate $z$, the K\"ahler potential and the corresponding
metric read
\be
K (z, {\bar z}) ~=~ r^2 \ln \left(1 + \frac{z {\bar z}}{r^2} \right) ~~~,~~~
g_{z {\bar z}} (z, \bar z) ~=~  \left(1 + \frac{z {\bar z} }{r^2}  
\right)^{-2} ~~~,
\label{s2pot}
\ee
with $1/r^2$ being proportional to the scalar curvature of the manifold.  It can 
be easily checked that a solution to (\ref{int}) is given by
\be
\U_\star (w) ~=~ \frac{\F (1 + \F {\bar \F}/r^2) + w \S  }
{1 + \F {\bar \F}/r^2 - w {\bar \F} \S /r^2 }~~~.
\label{s2soln}
\ee
All the auxiliary superfields are now determined in terms of the physical 
ones, $\F$ and $\S$. At first sight, $\U_\star (w)$ appears to develop an
unexpected simple pole at
\be
w ~=~ \frac{r^2}{{\bar \F} \S} (1 + \F {\bar \F}/r^2)~~~.
\label{falsepole}
\ee
However, this is misleading. The point is that $\U_\star (w)$ passes through 
the North Pole of the two-sphere at $w$ given as in (\ref{falsepole});  but we
must replace the coordinate system chosen by another one in a vicinity of the
North Pole.

With the use of (\ref{s2soln}) we find
\bea
K \left( \U_\star (w), \breve{\U}_\star (w) \right) \,&=& \, r^2 \ln 
\left\{ \Big( 1 + \F {\bar \F}/r^2 \Big) \Big( 1 - \frac{1}{r^2} \frac{ 
\S {\bar \S} } { ( 1 + \F {\bar \F}/r^2 )^2 }  \Big) \right\} \non \\
&{~}& +~\J \big( \U_\star (w) \big) ~+~ {\bar \J} \big( \breve{\U}_\star (w) \big)~~~,
\eea
where
$$
\J \big( \U_\star (w) \big) ~=~ - r^2 \ln \left(
1 - w  \frac{ {\bar \F} \S}{r^2 ( 1 + \F {\bar \F}/r^2 )} \right)~~~.
$$
Therefore, the action (\ref{nact}) turns into
\be
{\cal S}[\U_\star, \breve{\U}_\star] ~=~ \int {\rm d}^8 z \, \left\{ ~
K(\F, \bar 
\F) ~+ ~ r^2 \ln \Big(1 - \frac{1}{r^2} \, g_{\F {\bar \F}} (\F, {\bar \F})\;
\S \bar \S \Big) \right\} ~~~.
\label{physpol}
\ee
The action is real under the following global restriction
\be
r^2 ~~>~~ g_{\F {\bar \F}} (\F, {\bar \F})\; \S \bar \S  ~~~,
\label{domain}
\ee
which constitutes the upper bound for admissible values of $\S$. Under the 
above restriction, the second term in (\ref{physpol}) can be expanded 
in a power series in $\S \bar \S$ which is in fact an expansion in powers 
of the curvature.

Let us comment upon how the solution (\ref{s2soln}) has been obtained.
Keeping in mind the explicit structure of the K\"ahler potential
(\ref{s2pot}), one readily observes that the particular curve
\be
\U_0 (w) ~=~ \S w~~~,~~~ \breve{\U}_0 (w) ~=~ - \frac{\bar{\S}}{w}
\label{zerosol}
\ee
solves the equations (\ref{int}). A specific feature of this curve 
is that $\U_0(w=0)=0$. Further, one has to take into account two more
observations: (i) 
${\mathbb C}{\rm P}^1$ is a homogeneous space of SU(2);
(ii) the K\"ahler metric (\ref{s2pot}) is invariant with respect to
the SU(2) transformations. A simple consequence of (ii) is that
the equations (\ref{int}) are SU(2) invariant. Now, to construct
a curve $\U_\star (w)$ under the boundary condition $\U_\star (0) =\F$,
it is sufficient to apply a special
SU(2) transformation to the curve $\U_0(w)$.

The technique of elimination of the auxiliary superfields, which
we have just described, remains effective for a large class of K\"ahler
manifolds. It perfectly works if the following conditions are
satisfied:

(i) the K\"ahler potential can be chosen to be $R$-invariant
\be
K( {\rm e}^{{\rm i} \varphi } \F^I, {\rm e}^{-{\rm i }\varphi} {\bar 
\F}^{\bar{J}}) ~=~ K( \F^I,  {\bar \F}^{\bar{J}}) 
~~~;
\ee

(ii) the K\"ahler manifold is a homogeneous space of a Lie group G;

(iii) the group G leaves the K\"ahler metric invariant. \\
Nontrivial examples of such K\"ahler manifolds are the complex
projective spaces
${\mathbb C}{\rm P}^n$ and the complex Grassmann 
spaces ${\rm G}_{m,n}$. Under the above conditions, the particular 
curve (\ref{zerosol}) solves the equations (\ref{int}); to obtain  
$\U_\star^I (w)$ under the boundary condition $\U_\star^I (0) =
\F^I$, it remains to apply a special group transformation to
$\U_0(w)$.

For arbitrary initial conditions (\ref{geo3}), the corresponding
complex curve (\ref{s2soln}) turns out to be a solution of the 
holomorphic geodesic equation
\be
\frac{ {\rm d}^2 \U^I_\star (w) }{ {\rm d} w^2 } \,+ \,
\G^I_{JK} \left( \U_\star (w), \bar{\F} \right)
\frac{ {\rm d} \U^J_\star (w) }{ {\rm d} w }
\frac{ {\rm d} \U^K_\star (w) }{ {\rm d} w }  ~=~ 0 ~~~,
\label{geo1}
\ee
The above geodesic equation can be equivalently rewritten as follows
\be
\frac{ {\rm d}^2 }{ {\rm d} w^2 } K_{\bar{I}} \left( \U_\star (w),
\bar{\F} \right) ~=~ 0 ~~~.
\label{geod1}
\ee
This is the equation which eliminates the auxiliary superfields in case of the 
manifolds
${\mathbb C}{\rm P}^n$ and ${\rm G}_{m,n}$. In general, eq. (\ref{geo1}) is
not equivalent to the original  equations of motion for the auxiliary superfields
(\ref{int}).

In accordance with (\ref{domain}) , the physical variables $(\F, \S)$ of the $N=2$ CNM model
(\ref{physpol}) parametrize  an open domain of the tangent bundle to the
K\"ahler manifold. The nonminimal sector can be dualized into a 
chiral model if we replace the action
\be
{\cal S}[\S, \bar \S ] ~=~ r^2 \int {\rm d}^8 z \, \ln \Big(1 -
\frac{1}{r^2 } \,g_{\F {\bar \F}} (\F, {\bar \F}) \S \bar \S \Big)
\label{linaction}
\ee
by the following one
\be
{\cal S}[U, \bar U, \J, \bar \J ] ~=~ \int {\rm d}^8 z \,\left\{
r^2 \ln \Big(1 - \frac{1}{r^2} \, g_{\F {\bar \F}} (\F, 
{\bar \F}) U \bar U \Big) + U\J + \bar U \bar \J \right\}~~~.
\label{auxiliaryaction}
\ee
Here $U$ is a complex unconstrained superfield, and $\J$ is a chiral
superfield. It is worth pointing out that $U$ is to be treated as a 
tangent vector to the point $\F$ of the K\"ahler manifold, while $\J$ 
as a one-form.  If we vary the latter action with respect to $\J$,
$U$ becomes a linear superfield, and we return to the nonminimal action
(\ref{linaction}). On the other hand, we can eliminate $U$ with the aid
of its equation of motion
\be
\frac{g_{\F \bar{\F}}(\F, {\bar \F}) \bar U }
{1 - g_{\F \bar{\F}}(\F, {\bar \F}) U \bar U / r^2}
~=~ \J ~~~,
\ee
and this results in a purely chiral model.  When $U$ spans the open ball 
(\ref{domain}), $\J$ spans the whole cotangent space to the point $\F$. 
Therefore, the complex dynamical variables $(\F, \J)$ parametrize the cotangent
bundle to
${\mathbb C}{\rm P}^1$,  which is known to be a hyper-K\"ahler manifold
\cite{PER}.

Upon implementing this duality transformation, the action (\ref{physpol}) turns
into
\be
{\cal S}[\F, \bar \F, \J, \bar \J] ~=~ \int {\rm d}^8 z \, \left\{
K(\F, \bar \F) ~+ ~ \cF (\F, \bar \F, \J, \bar \J)
\right\}
\label{chiralaction}
\ee
where
\be
 \cF (\F,\bar \F, \J, \bar \J) ~=~ -r^2 \ln \Big(f(\k)\Big) + 2r^2
\frac{\k}{f(\k)}~~~, ~~~ f (\k) = \hf\left(1 +\sqrt{1 +4\k}\right)
\ee
and
\be
 \k =\frac{1}{r^2} g^{\F {\bar \F}} (\F, {\bar \F}) \J \bar \J ~~~.
\ee
The lagrangian in (\ref{chiralaction}) is the hyper-K\"ahler potential of the 
cotangent bundle to  
${\mathbb C}{\rm P}^1$.  One can easily check the identity
\be
\cF'(\k) ~=~ \frac{r^2}{2\k} \left(-1 +\sqrt{1 + 4\k}\right)
\ee
which is in complete agreement with ref. \cite{PER}.

\subsection{General Structure of  ${\cal S}_{\s}[\U_{\star},
\breve{\U}_{\star}]$} 
~~~~The results of subsections 2.2 and 2.3 yield some
insights into  the structure  of ${\cal S}_{\s}[\U_{\star},
\breve{\U}_{\star}]$ in the general case.  
This action, in general, turns out to 
be an infinite power series in covariant derivatives of the curvature
tensor  of the K\" ahler manifold. 

If we set ${\cal A} ~=~ 0$ in (\ref{nact2}), we obtain
\bea
{\cal S}_{\s}[\U_{\star}, \breve{\U}_{\star}] 
&\approx & \int {\rm d}^8 z \, \Big\{ \,\, K \big( \F, \bar{\F} \big) ~-~
g_{I \bar{J}} \big( \F, \bar{\F} \big) \S^I {\bar \S}^{\bar{J}} 
~+~ \fracm{1}{4} K_{,I ,\bar{K} ,J ,\bar{L} } \S^I \S^J
\bar{\S}^{\bar{K}} \bar{\S}^{\bar{L}} {~~~~~~~~~~} \non \\
&{~}& ~~~~~~~~-~ \, \sum_{p=0}^{\infty} \,(-1)^p \Big[ \fracm 1{(p + 3)!} 
{\Big]}^2 \,(\prod_{\ell=1}^{p +3} \S^{I_{\ell}} \,\bar{\S}^{\bar{K}_{\ell}} )  
K_{,I_1 ... ,I_{p+3},{\bar K}_1 \dots  ,{\bar K}_{p+3} }~ 
\Big\}~~~. 
\label{actn3}
\eea
The first two terms above have the characteristic form of the 4D, $N$ = 1
CNM nonlinear $\s$-model \cite{G1}.  For the terms quartic in $\S$ we find
\be
\fracm{1}{4} \, [ \, {\pa}_I {\pa}_{\bar K} {\pa}_J {\pa}_{\bar{L} } K 
 \, ] ~ \S^I \S^J \bar{\S}^{\bar{K}} \bar{\S}^{\bar{L}} ~~~. \label{fortm}
\ee
This is proportional to the ``dominant term'' in the full curvature tensor 
for a K\" ahler manifold.
\be
R_{I \bar{K} J \bar{L} } ~=~ K_{,I ,{\bar K} ,J ,{\bar L} } ~-~ g^{ M {\bar 
N}} K_{,I ,J , {\bar N} } \, K_{, {\bar K}  ,{\bar L} ,{M}} ~~~.
\ee
It must be the role of the infinite number of auxiliary superfields 
to covariantize the dominant terms so that these terms above are replaced 
according to
\bea
K_{,I_1 ,I_{2},{\bar K}_1 \dots ,{\bar K}_{2} } ~&\to& ~
R_{I_1  {\bar K}_1 I_{2} {\bar K}_{2} } ~~~, \non \\
K_{,I_1 ... ,I_{p+3},{\bar K}_1 \dots  ,{\bar K}_{p+3} }
~&\to& ~ (\prod_{\ell=1}^{p +1} \nabla_{I_{\ell}} )  
(\prod_{\ell'=1}^{p +1}     {\Bar 
\nabla}{}_{{\bar K}_{\ell'}} ) \, R_{I_{p+2} \bar{K}_{p+2} I_{p+3} 
\bar{K}_{p+3} } ~~~. \label{covr}
\eea
Above in (\ref{covr}), the operators $\nabla_{I}$ and ${\Bar \nabla}
{}_{{\bar K}}$ denote covariant derivatives with respect to the K\" ahler
metric $g_{I \bar{J}}$. We emphasize that at present, we have no
proof of this proposal.  However, this is suggested by demanding 
covariance with respect to the K\" ahler geometry of the model. As
well this is also suggested by the role of the auxiliary field in
ordinary 4D, N = 1 supersymmetric nonlinear $\s$-models.

However, the results in (\ref{physpol}) imply that the results in 
(\ref{actn3}) and (\ref{covr}) cannot be the complete answer for the
removal of the auxiliary fields, for the structures in the second line 
of (\ref{covr})  vanish for covariantly curvature. The obvious reason 
that we know this is because our second explicit example in (\ref{physpol})
describes a K\" ahler metric with covariantly constant curvature. The 
second term in (\ref{physpol}) may be expanded in $\S {\bar \S}$ and
contains terms to all orders; this is in fact an expansion in powers of 
curvature.  It must be the case that the correct expression for the action 
takes the form
\bea
{\cal S}_{\s}[\U_{\star}, \breve{\U}_{\star}] 
&=& \int {\rm d}^8 z \, \Big\{ \,\, K \big( \F, \bar{\F} \big) ~-~
g_{I \bar{J}} \big( \F, \bar{\F} \big) \S^I {\bar \S}^{\bar{J}} 
~+~ \fracm{1}{4} R_{I \bar{K} J \bar{L} } \S^I \S^J
\bar{\S}^{\bar{K}} \bar{\S}^{\bar{L}} {~~~~~~~~~~} \non \\
&{~}& ~~~~~~~~-~ \, \sum_{p=0}^{\infty} \,(-1)^p \Big[ \fracm 1{(p + 3)!} 
{\Big]}^2 \,(\prod_{\ell=1}^{p +3} \S^{I_{\ell}} \,\bar{\S}^{\bar{K}_{\ell}} )  
{\cal G}_{I_1 ... I_{p+3}{\bar K}_1 ...  {\bar K}_{p+3} }~ 
\Big\}~~~, \non \\
&{~}&{\cal G}_{I_1 ... I_{p+3}{\bar K}_1 ... {\bar K}_{p+3} }~=~
{\cal G}_{I_1 ... I_{p+3}{\bar K}_1 ... {\bar K}_{p+3} }(
R_{I \bar{K} J \bar{L} }, \, \nabla_{M} , \, {\Bar \nabla}
{}_{{\bar M}}) ~~~.
\label{actn6}
\eea
The terms on the last line of (\ref{actn3}) are simply the leading
terms in the (yet to be completely determined) functions ${\cal G}_{
I_1 ...I_{p+3}{\bar K}_1 ... {\bar K}_{p+3} }$.  The perturbative
description of these is provided by the sequence of steps in (\ref{spec} 
- \ref{ase3}). It is worth noting that each term in the action 
${\cal S}_{\s}[\U_{\star}, \breve{\U}_{\star}]$ contains an equal
number  of $\S$- and $\bar \S$-multiplets, as a consequence of the U(1) 
symmetry (\ref{rfiber}).

The naive dimensional compactification of our 4D, $N$ = 2 results leads to
new CNM formulations of 2D, $N$ = 4 theories. Similar theories have recently
been discussed by Penati et.~al.~\cite{PRSZ}.  One result shown is that
CNM models possess a natural encoding of the three complex structures 
one expects of a hyper-K\" ahler model.  Although their 2D, $N$ = 4 models
do not describe unitary quantum field theories, the encoding of the
complex structures they describe remains true in our construction as well
which in addition is ghost-free. The 2D, $N$ = 4 form of our results
follow by applying the toroidal compactification to our 4D models with
$N$ = 2 supersymmetry.

\sect{Special Hyper-K\" ahler Geometries}

~~~~Our approach to 4D, $N$ = 2 supersymmetrical non-linear $\s$-models
yields an interesting view point on issues surrounding hyper-K\" ahler
manifolds. Metrics describing such manifolds are in general difficult 
to explicitly construct. Remarkably, $N=2$ supersymmetry provides us with 
a regular procedure to construct hyper-K\"ahler structures associated with 
a family of hyper-K\"ahler manifolds -- the cotangent bundles to K\"ahler
manifolds.  In fact, it was conjectured for a long time that every total 
space $T^\ast \cM$ of the cotangent bundle to a K\"ahler manifold $\cM$ 
admits a hyper-K\"ahler structure\footnote{We are grateful to Andrei Lossev 
for enlightening discussions on  this conjecture.}, and a lot of relevant 
examples were elaborated, see, for instance, \cite{PER}. This conjecture 
has been recently proved by Kaledin \cite{KAL}, as an existence theorem, 
for an open neighborhood $U \subset T^\ast \cM$ of the zero section $\cM 
\subset T^\ast \cM$. Using the supersymmetric model (\ref{nact4}) allows 
us to explicitly construct the hyper-K\"ahler metrics on  $T^\ast \cM$.

Upon elimination of the auxiliary superfields, the model (\ref{nact4})
turns into $N=2$ supersymmetric CNM $\sigma$-model (\ref{actn6}),
\be 
{\cal S}_{\s}[\U_{\star}, \breve{\U}_{\star}] ~=~ {\cal S}_{\s}[\F, \bar \F, 
\S, \bar \S] ~=~  \int {\rm d}^8 z \,  \cL_{{\rm CNM}} (\F^I, {\bar \F}^{\bar
J}, \S^K, {\bar \S}^{\bar L})
\label{cnmmodel}
\ee
which possesses a nonlinearly realized $N=2$ supersymmetry and is defined on
an open neighborhood  $U \subset T \cM$ (of the zero section $\cM \subset T
\cM$) of the tangent bundle $T\cM$ to the K\"ahler manifold $\cM$. The 
tangent-space variables are realized in terms of the complex linear superfield
$\S$ and its conjugate.  In the case of ${\mathbb C}{\rm P}^1$, the domain $U$
is defined by eq. (\ref{domain}). The action (\ref{cnmmodel}) can be replaced
by a dually equivalent one with the action
\be
 {\cal S}_{\s}[\F, \bar \F, G, \bar G, \J, \bar \J] ~=~
 \int {\rm d}^8 z \,  \left\{ \cL_{{\rm CNM}} (\F^I, {\bar \F}^{\bar J}, G^K, {\bar G}^{\bar L})
+ G^K \J_K + {\bar G}^{\bar L} {\bar \J}_{\bar L} \right\}~~~,
\label{auxmodel}
\ee 
where $G$'s are  complex unconstrained superfields, and $\J$'s are chiral
superfield. It is obvious that $G^I$ should be treated as a tangent vector,
while $\J_I$ as a one-form to the point $\F$ of $\cM$. The auxiliary
superfields $G$ and $\bar G$ can be eliminated with the aid of their equations
of motion. Then, we stay with a purely chiral-chiral model
\be
{\cal S}_{\s}[\F, \bar \F, \J, \bar \J] ~=~ \int {\rm d}^8 z \, \cL_{{\rm
CC}} (\F^I, {\bar \F}^{\bar J}, \J_K, {\bar \J}_{\bar L})
\label{ccmodel}
\ee
which possesses a nonlinearly realized $N=2$ supersymmetry and is defined on
an open submanifold  $V \subset T^\ast \cM$ of the cotangent bundle $T^\ast
\cM$ to the K\"ahler manifold $\cM$. Now, the most important point is that
$\cL_{{\rm CC}} (\F, {\bar \F}, \J, {\bar \J})$ is nothing else but  the
hyper-K\"ahler potential on $T^\ast \cM$!

Our consideration shows how to construct hyper-K\"ahler metrics on
cotangent bundles to K\"ahler manifolds. The procedure is as follows:
\begin{itemize}
\item given a K\"ahler manifold $\cM$, construct 
the $N=2 $ models (\ref{nact4}) with \newline linearly realized 
supersymmetry;
\item  eliminate the infinite number of
auxiliary superfields;
\item  fulfill the duality transform described in this section.
\end{itemize}
This is a long but constructive way to generate hyper-K\"ahler metrics.
The above construction defines a special family of hyper-K\"ahler metrics
and, hence,  ``special hyper-K\" ahler geometries'' \footnote{We
are inspired to use this name in analogy  with ``special K\" ahler
geometries'' associated the KVM \newline \indent ${~~}$ in \cite{KVM}.}.

In accordance with the results of \cite{KAL}, the hyper-K\"ahler metric on $T^\ast \cM$
is invariant under the action of the group U(1) on $T^\ast \cM$ given by 
dilatations along the fibers of the canonical projection $\pi ~:~ T^\ast \cM \to \cM$.
This invariance group is nothing else but the U(1) fiber symmetry (\ref{rfiber}) of 
our $N=2$ $\sigma$-model (\ref{nact}).

\sect{4D Soft Pion Physics, $N$ = 2 Supersymmetry and Holomorphic 
Higher Derivative Actions}

~~~~In previous works \cite{G1}, the use of well-known aspects of chiral
perturbation theory \cite{SM} was advocated for the  construction of a
``phenomenologically based'' 4D, $N$ = 1 CNM low-energy  QCD supersymmetric
effective action. The advent of the ``polar multiplet formalism'' now permits
a further $N$ = 2 extension of these ideas.

In the final work of \cite{G1}, it was shown that any higher derivative
term which occurs in a non-supersymmetric action of the type described 
in chiral perturbation theory can be mapped into a 4D, $N$ = 1 superfield 
action under the action of an operator denoted by ${\cal G}_{\rm S}$. 
After an integration-by-parts to form actions integrated over ${\rm d}^8 
z$ such superfield actions take forms very similar to that of (\ref{HD}). 
However, (\ref{HD}) does not produce higher derivative terms. A way 
to produce such terms within the polar multiplet formalism seems to begin 
with the following observation. The action
\beq
{\cal S}[\U, \breve{\U}] ~=~ \frac{1}{2\pi {\rm i}} \oint \frac{{\rm
d}w}{w}
\int {\rm d}^{8} z \; {\cal H} \Big( \U , \pa_{\un a} \U,..., 
\bar{\U}, \pa_{\un a} \bar{\U},..., w \, \Big)
\eeq
necessarily generates higher derivative terms. Combining this with our
previous suggestion seems to have the following implication. Since we
believe that holomorphy is of critical importance for the higher
derivative terms, it is only a change
\beq
{\cal J} \big( \U , \, \breve{w}) ~\to ~ {\cal J} \big( \U ,\, 
\pa_{\un a} \U , \, \breve{w}\,  \big)  ~~~,
\label{RPL} \eeq
in (\ref{HD}) that will generate holomorphic higher derivative terms
consistent with an $N$ = 2 supersymmetry. In particular, we believe the first
term  in (\ref{HD}) should not be modified. However, there must also be
addition changes that correspond to inserting the factors of $D_{\a}$ and
${\Bar D}{}_{\dot \a}$ which occurred in the $N$ = 1 case and. 
Such higher derivative terms can come from a manifestly $N=2$ supersymmetric
functional of the general form
\be
{\cal S}[\U, \breve{\U}] ~=~ \int {\rm d}^{12} z \; {\cal G} \Big( \U_m , 
{\bar \U}_n \Big)~~~, ~~~m,n= 1,2, \ldots ~~~,
\ee
where the integration is carried out over the full $N=2$ superspace. 
This topic is still under study.

Since manifest realization of $N$ = 2 supersymmetry forces the use 
of the CNM doublet, this provides an ``$N$ = 2 answer'' to yet 
another seemingly ad hoc feature of the $N$ = 1 CNM low-energy QCD 
supersymmetric model. In our present work, the pion octet must 
be embedded in the superfield pair $(\Phi^I, \, \S^I)$ as a 
consequence of $N$ = 2 supersymmetry.  In turn the {\it a}
{\it {priori}} existence of the mixing angle $\g_{\rm S}$ \cite{G1}, 
which expresses the possibility that the pion octet can occur 
as a linear combination of the two $0^-$ octets that appear 
in the CNM doublet, is seen to have an  $N$ = 2 supersymmetry
origin.

\sect{Conclusion}

~~~~~~~~In a sense, the action in (\ref{nact}) together with (\ref{RPL}) 
completes several cycles of our past work.  Many years ago, we \cite{KVM} 
proposed a 4D, $N$ = 1 superfield action of the form
\beq
{\cal S}_{\rm {KVM}} ~=~ \Big\{ \, \int d^4 x \, \Big[ \, \int 
d^2 \q \, d^2 {\bar \q} ~ {\Bar \Phi}{}^{\, \rm I} \pa_{\rm I} 
~+~ \fracm 14 \int \, d^2 \q ~ W^{\a \, {\rm I}} \,  W_{\a}^{ \, 
{\rm K}}\, \,  \pa_{\rm I} \pa_{\rm K}\, \Big] \, H(\Phi) \,+\, 
{\rm {h.\, c.}} \, \Big\} ~.~~
\eeq
to describe a 4D, $N$ = 2 supersymmetric $\s$-model which we named the 
``K\" ahlerian Vector Multiplet'' model. With (\ref{nact}) and the 
subsequent elimination of the auxiliary superfields, we have succeeded 
in writing a purely 4D, $N$ = 1 supersymmetric $\s$-model action for 
the hypermultiplet. In the intervening years, of course, there has been 
the much heralded work of Seiberg and Witten \cite{SW} on $N$ = 2 
supersymmetrical Yang-Mills theory that completely determines the 
function $ H(\Phi)$ above. Two years ago \cite{G1}, we renewed our
study of CNM models because they offered the opportunity to clearly
inject  geometrical aspects of real-world chiral perturbation theory
\cite{SM} into the hypothetical world of 4D, $N$ = 1 (and now likely $N$
= 2)  supersymmetry. The results in (\ref{nact}) and (\ref{RPL}) hold out
the  prospect that these two worlds may not be as far removed as might
have  first seemed. \newline
${~~~}$ \newline
``{\it {Nullius in verba.}}'' -- Motto of the Royal Society of London
$${~~~}$$
\noindent
{\bf {Acknowledgment}} \newline \noindent
${~~~~}$Both authors wish to acknowledge very helpful exchanges 
with I. Buchbinder, B. de Wit, N. Dragon, F. Gonzales-Rey, T. H\"
ubsch, E. Ivanov,  U. Lindstr\"om, D.  L\"ust, B. Ovrut, M. Ro\v cek, S.
Theisen and M. Vasiliev.  We thank especially  A.~ Lossev for informing 
us about Ref. \cite{KAL} and A.~Nersessian  for bringing Ref. \cite{PER}
to our attention. One of us (S.K.)  acknowledges the Bogoliubov
Laboratory of Theoretical Physics at the  Joint Institute for Nuclear
Research, where part of this project was done,  for hospitality. The work
of S.K. was supported in part from RFBR grant, project No  96-02-16017;
RFBR-DFG grant, project No 96-02-00180; INTAS  grant, INTAS-96-0308.


\begin{appendix}

\sect{Projective Superspace}

~~~~Superfields living in the $N=2$ projective superspace \cite{KLR}
are parametrized by a complex bosonic variable $w$ along with
the coordinates of $N=2$ global superspace 
${\mathbb R}^{4|8}$
\be
z^M~=~(x^m,\theta_i^\alpha,
\bar \theta^i_{\dot{\alpha}}) ~~~,~~~ \overline{\theta_i^\alpha} ~=~ \bar
\theta^{{\dot{\alpha}}\,i} ~~~,~~~ i=1,2~~~.
\ee
A superfield of the general form
\be
\Xi (z, w) ~=~ \sum_{n=-\infty}^{+\infty} \Xi_n (z) w^n  
\label{generalprjsup}
\ee
is said to be projective if it satisfies the constraints
\be
\nabla_\a (w) \Xi (z, w) ~=~ 0 ~~~,~~~ {\Bar \nabla}_\ad (w)\Xi (z, w)~=~ 0
\label{pc1}
\ee
which involve the operators
\be
\nabla_\a (w) ~\equiv~  w D^1_\a \,-\, D^2_\a ~~~,~~~
{\Bar \nabla}_\ad (w) ~\equiv~ {\Bar D}_{\ad 1} \,+\, w {\Bar D}_{\ad 2}
\label{nabla}
\ee
constructed from the $N=2$ covariant derivatives $D_M \,=\, (\pa_m , 
D_\a^i , {\Bar D}{}^\ad_i)$. The operators $\nabla_\a (w)$ and ${\Bar 
\nabla}_\ad (w)$ strictly anticommute with each other, as a 
consequence of the covariant derivative algebra
\be
\{ D^i_\a , D^j_\b\}~=~ \{ {\Bar D}_{\ad i}, {\Bar D}_{\bd j} \}~=~ 0 ~~~,~~~
\{ D^i_\a , {\Bar D}_{\ad j} \}~=~ - 2 {\rm i} \, \d^i_j \, \pa_{\a \ad}~~~.
\label{cda}
\ee
With respect to the inner complex variable $w$, the projective
superfields are holomorphic functions on the punctured complex plane 
${\mathbb C}\,{}^*$
\be
\pa_{\bar w} \, \Xi (z, w) ~=~ 0~~~.
\label{pholomorphy}
\ee

Constraints (\ref{pc1}) rewritten in components
\be
D^2_\a \Xi_n~=~ D^1_\a \Xi_{n-1} ~~~,~~~ {\Bar D}_{\ad 2} \Xi_n 
~=~ - \, {\Bar D}_{\ad 1} \Xi_{n+1}
\label{pc2}
\ee
determine the dependence of the component $N=2$ superfields
$\Xi$'s on $\q^\a_2$
and ${\bar \q}^2_\ad$ in terms of their dependence on $\q^\a_1$
and ${\bar \q}^1_\ad$. Therefore, the components $\Xi_n$ are
effectively superfields over the $N=1$ superspace parametrized by
\be
\q^\a ~=~ \q^\a_1 \quad , \qquad {\bar \q}_\ad ~=~ {\bar \q}^1_\ad ~~~.
\ee
If the power series in (\ref{generalprjsup}) terminates, some of the 
$N=1$ superfields satisfy constraints involving the $N=1$ covariant
derivatives
\be
D_\a ~=~ D^1_\a \qquad \quad {\Bar D}^\ad ~=~ {\Bar D}^\ad_1 ~~~.
\ee

A natural operation of conjugation, which brings every projective
superfield into a projective one, reads as follows
\be
\breve{\Xi} (z,w) ~=~ \sum_{n} (-1)^n {\bar \Xi}_{-n} (z) w^{n}
\label{pconjugation}
\ee
with ${\bar \Xi}_n$ being the complex conjugate of $\Xi_n$.
A real projective superfield is constrained by
\be
\breve{\Xi} ~=~ \Xi \quad \Longleftrightarrow \quad
{\bar \Xi}_n ~=~ (-1)^n \Xi_{-n}~~~.
\ee
The component $\Xi_0(z)$ is seen to be real. The following equations are
also seen to be valid,
\be
\breve{\nabla}{}_\a (w) ~=~ - \fracm 1w {\Bar \nabla}_\ad (w)~~~,
~~~ \breve{{\Bar \nabla}}{}_\ad (w) ~=~  \fracm 1w {\nabla}_\a (w)~~~,
\ee
so that application of the ${~}_{\breve {~}}$-operation twice is
equivalent to the identity map on these operators.  It follows the
operators,
\be
{\tilde {\nabla}}{}_\a (w) ~\equiv ~ (i)^{\fracm 32} \, (\fracm 1{w})^{\fracm
12}\, {\nabla}{}_\a (w) ~~~,~~~ {\tilde {{\Bar \nabla}}}{}_\ad (w) ~\equiv ~ 
-\, (i)^{\fracm 12} \, (\fracm 1w)^{\fracm 12} \,{\Bar \nabla}_\ad
(w)~~~, 
\ee
transform as 
\be
{\tilde {\nabla}}{}_\a (w) ~ \to ~ {\tilde {{\Bar \nabla}}} {}_\ad (w) 
~~~,~~~ {\tilde {{\Bar \nabla}}}{}_\ad (w) ~\to~ {\tilde {\nabla}}{}_\a (w)
~~~,~~~
\ee
under ${~}_{\breve {~}}$-conjugation.

Given a real projective superfield $\cL(z, w)$, $\breve{\cL} = \cL$,
we can construct a $N=2$ supersymmetric invariant by the following rule
\be
{\cal S} ~=~ \int {\rm d}^4 x\, D^4 \cL_0 (z) | ~~~ , ~~~
D^4 ~=~ \frac{1}{16} D^{\a 1} D^1_\a {\Bar D}_{\ad 1}  {\Bar D}^\ad_1
~~~, \ee
where $D^4$ is the $N=1$ superspace measure, $\cL_0 (z) = \cL (z, \, 0) 
$ and $U|$ means the $\q$-independent component of a superfield $U$.
Actually, from the standard supersymmetric transformation law
\be
\d \cL ~=~ {\rm i} \left(\ve^\a_i Q^i_\a \, + \,{\bar \ve}^i_\ad 
{\Bar Q}^\ad_i \right)  \cL
\ee
we find
\bea
\d {\cal S} &=& {\rm i} \int {\rm d}^4 x\, \left(\ve^\a_i Q^i_\a  \,+ \,
{\bar \ve}^i_\ad {\Bar Q}^\ad_i \right) D^4 \cL_0 ~=~
- \int {\rm d}^4 x\, \left(\ve^\a_i D^i_\a \, + \,
{\bar \ve}^i_\ad {\Bar D}^\ad_i \right) D^4 \cL_0 | \non \\
&=& - \int {\rm d}^4 x\, \left(\ve^\a_2 D^2_\a \, + \,
{\bar \ve}^2_\ad {\Bar D}^\ad_2 \right) D^4 \cL_0 |~=~
- \int {\rm d}^4 x\, D^4 \left(\ve^\a_2 D^2_\a  \, + \,
{\bar \ve}^2_\ad {\Bar D}^\ad_2 \right) \cL_0 |   \non \\
&=& - \int {\rm d}^4 x\, D^4 \left(\ve^\a_2 D^1_\a  \cL_{1} |
\, - \, {\bar \ve}^2_\ad {\Bar D}^\ad_1 \cL_1 | \right)~=~ 0~~~.
\non
\eea
The action can be rewritten in the form \cite{KLR}
\be
{\cal S} ~=~ \frac{1}{2\pi {\rm i}} \oint_{C} \frac{{\rm d}w}{w}
\int{\rm d}^4 x\, D^4 \cL | ~~~,
\label{praction2}
\ee
where $C$ is a contour around the origin.

The only projective superspace representation considered in the main body of this paper is
the so-called polar multiplet (or $\U$ hypermultiplet) which describes a charged
$N=2$ scalar multiplet and is defined by eq. (\ref{exp}).  In Appendix B we also make use 
of a complex tropical multiplet.

\sect{A First Look at Poincar\' e Duality and Projective Superspace}

~~~~Since the concept of duality is presently considered of importance,
in this appendix we discuss the implementation of Poincar\' e Duality
within the context of polar multiplets.  The hypermultiplet model
\bea
S[\U, \breve{\U}] \, &=& \, \frac{1}{2\pi  {\rm i}} \, \oint 
\frac{{\rm d}w}{w} \,  \int {\rm d}^8 z  \,   \breve{\U} \U   ~~~.
\label{nactt} \eea
with  
\bea
\U &=& \sum_{n=0}^{\infty}  \, \U_n (z) w^n ~=~  \F(z) \, +\,  w \S(z)
 \,+\, O(w^2) ~~~,\non \\
\breve{\U}{}  &=& \sum_{n=0}^{\infty}  (-1)^n \, {\bar \U}_n
(z) w^{-n} ~=~ {\bar \F}{}(z)  - \fracm 1{w}
{\bar \S}{}(z)  + O((\frac{1}{w})^2)~~~,
\label{exp1}
\eea
describes a free CNM doublet. The chiral scalar $\F$ can be dualized into a
complex linear  one, and the complex linear scalar $\S$ can be dualized into a
chiral one. As a result, we again obtain a CNM doublet. Therefore, the above
theory can be said to be self-dual, in the sense of Poincar\'e duality. Below
we will develop the relevant formalism to make this manifest.

First, let us introduce a {\it complex tropical} multiplet 
\be
U = \sum_{n=-\infty}^{\infty}  \, U_n (z) w^n  ~~~,~~~
\breve{U}{}  = \sum_{n=-\infty}^{\infty}  (-1)^n \, {\bar U}_n
(z) w^{-n} ~~~,
\label{trop}
\ee
with all the component $U_n$ being complex unconstrained superfields.
We also introduce a multiplet realized as follows
\bea
\Xi &=& \sum_{n=-1}^{\infty}  \, \Xi_{n} (z) w^n ~=~ 
 \frac{1}{w}\J(z)  ~+~  \G(z) ~ +~ O(w) ~~~,\non \\
\breve{\Xi}{}  &=& \sum_{n=-1}^{\infty}  (-1)^n \, {\bar \Xi}_{n}
(z) w^{-n} ~=~ -~ w{\bar \J}{}(z)  ~+~
{\bar \G}{}(z)    ~+~ O(\frac{1}{w})~~~,
\eea
with $\J$ being chiral, $\G$ being complex linear, and the rest component 
superfields being complex unconstrained. It is worth keeping in mind the difference between the definitions 
of $\U $ and $\Xi$.

Consider the following auxiliary action
\bea
S \,& = &\, \frac{1}{2\pi  {\rm i}} \, \oint \frac{{\rm d}w}{w} \,  \int {\rm
d}^8 z  \, \Big( ~\breve{U} U   ~-~  U \Xi ~-~  \breve{U}
\breve{\Xi} ~ \Big)
\non
\\ & = & \int {\rm d}^8 z  \, \Big( \sum_{n=-\infty}^{\infty}   ~
(-1)^n {\bar U}_n U_n  ~-~ 
\sum_{n=-1}^{\infty} \Xi_n U_{-n} ~-~ \sum_{n=-1}^{\infty}{\bar \Xi}_n {\bar U}_{-n} 
~\Big)
~~~.
\label{auxaction}
\eea
One easily observes
\be
\frac{\d S}{\d \Xi} ~=~ 0 \qquad \Longleftrightarrow  \qquad U(w) ~=~ \U(w) 
~=~ \sum_{n=0}^{\infty}  \, \U_n (z) w^n ~~~.
\ee
Therefore we return to the original model (\ref{nactt}). On the other hand, let 
us now consider the equation of motion for $U(w)$. From
\be
\frac{\d S}{\d U} ~=~ 0 
\ee
we obtain
\bea 
U_n & =&  0~~~, ~~~  n \geq   2 \non \\ 
 (-1)^n {\bar U}_n - \Xi _{-n} ~=~ 0~~~,~~~  (-1)^n U_n - {\bar \Xi}_{-n} 
& ~=~& 0 ~~~,~~~  n \leq 1~~~.
\eea
Therefore
\be
U(w) ~=~ \breve{\Xi} (w)~~~, ~~~\breve{U}(w) ~=~ \Xi (w) ~~~,
\ee
and the action takes the form
\bea
S[\Xi, \breve{\Xi}] \, &=& \, -\,  \frac{1}{2\pi  {\rm i}} \, \oint \frac{{\rm
d}w}{w} \,  \int {\rm d}^8 z  \, \,  \breve{\Xi} \, \Xi     ~~~.
\eea
We see that the theory (\ref{nact}) is self-dual.

In case of a general nonlinear sigma-model
\bea
S[\U, \breve{\U}] \, &=& \, 
\frac{1}{2\pi  {\rm i}} \, \oint \frac{{\rm d}w}{w} \,  \int {\rm d}^8 z  \,  
K\big( \breve{\U}, \U \big)   
\eea
we can also formally implement the duality transformation described above. But
now it replaces the potential $K( \breve{\U}, \U ) $ by a new one. The
requirement that the sigma-model be self-dual can be put forward and it restricts, in a
nontrivial way, the class of possible models. It is interesting to conjecture
that for $N=4$ SYM effective theories realized in projective superspace, the
hypermultiplet action must be self-dual with respect to the  duality
transformation above.

\end{appendix}

\end{document}